\documentclass[journal, a4paper, 10pt]{IEEEtran}
\usepackage{graphicx}
\usepackage{epsfig}
\usepackage{algorithm}
\usepackage{algorithmic}
\usepackage{ifmtarg}
\usepackage{cite}
\usepackage{amssymb}
\usepackage{amsthm}
\usepackage[fleqn]{amsmath}
\usepackage{amsmath}
\usepackage{color}
\usepackage{bbm}
\usepackage{cite}
\usepackage{epstopdf}
\usepackage[table,xcdraw]{xcolor}
\usepackage{flushend}

\usepackage{flushend}
\usepackage{subcaption}

\begin{document}
      
\title{Can Uplink Transmissions Survive in Full-duplex Cellular Environments? \\
}
\author{\IEEEauthorblockN{Hesham ElSawy, Ahmad AlAmmouri, Osama Amin, and Mohamed-Slim Alouini\\}
\IEEEauthorblockA{King Abdullah University of Science and Technology (KAUST), Computer, Electrical and Mathematical Science and Engineering Division (CEMSE), Thuwal 23955-6900, Saudi Arabia \\
Email: \{hesham.elsawy, ahmad.alammouri, osama.amin, slim.alouini\}@kaust.edu.sa}
\vspace{-1.75mm}}

\maketitle

\thispagestyle{empty}
\pagestyle{empty}

\begin{abstract}

In-band full-duplex (FD) communication is considered a potential candidate to be adopted by the fifth generation (5G) cellular networks.  FD communication renders the entire spectrum simultaneously accessible by uplink and downlink, and hence, is optimistically promoted to double the transmission rate. While this is true for a single communication link, cross-mode interference (i.e., interference between uplink and downlink) may diminish the full-duplexing gain. This paper studies FD operation in large-scale cellular networks with real base stations (BSs) locations and 3GPP propagation environment. The results show that the uplink is the bottleneck for FD operation due to the overwhelming cross-mode interference from BSs. Operating uplink and downlink on a common set of channels in an FD fashion improves the downlink rate but significantly degrades (over 1000-fold)  the uplink rate. Therefore, we propose the $\alpha$-duplex scheme to balance the tradeoff between the uplink and downlink rates via adjustable partial overlap between uplink and downlink channels. The $\alpha$-duplex scheme can provide a simultaneous $30\%$ improvement in each of the uplink and downlink rates. To this end, we discuss the backward compatibility of the $\alpha$-duplex scheme with half-duplex user-terminals. Finally, we point out future research directions for FD enabled cellular networks.


\end{abstract}

\section{Introduction}

In-band full-duplex communication (FD) is expected to provide several benefits to wireless networks when compared to the conventional half-duplex (HD) communication.\footnote{To avoid the overwhelming self-interference, HD transceivers orthogonalize transmission and reception in frequency domain via frequency division duplexing (FDD) or in time domain via time division duplexing (TDD).}  For instance, FD communication can improve spectrum utilization, improve physical layer security, reduce relaying latency, and enhance interference coordination~\cite{FD2}. FD communication emerges from recent advances in RF circuit design, which enables transceivers to sufficiently cancel their self-interference (SI) and simultaneously transmit and receive on the same channel~\cite{FD1,FD2}. It is believed that achieving sufficient SI cancellation would alleviate the necessity to orthogonalize transmission and reception, which renders the total amount of resources (time and frequency) simultaneously accessible by forward and reverse links. This would double the bandwidth available for each link and can provide up to 100$\%$ rate gains when compared to its HD counterpart~\cite{FD2}. 

   \begin{figure*}[t]
	\centering
	\begin{subfigure}[b]{0.32\textwidth}
	\includegraphics[width=\textwidth]{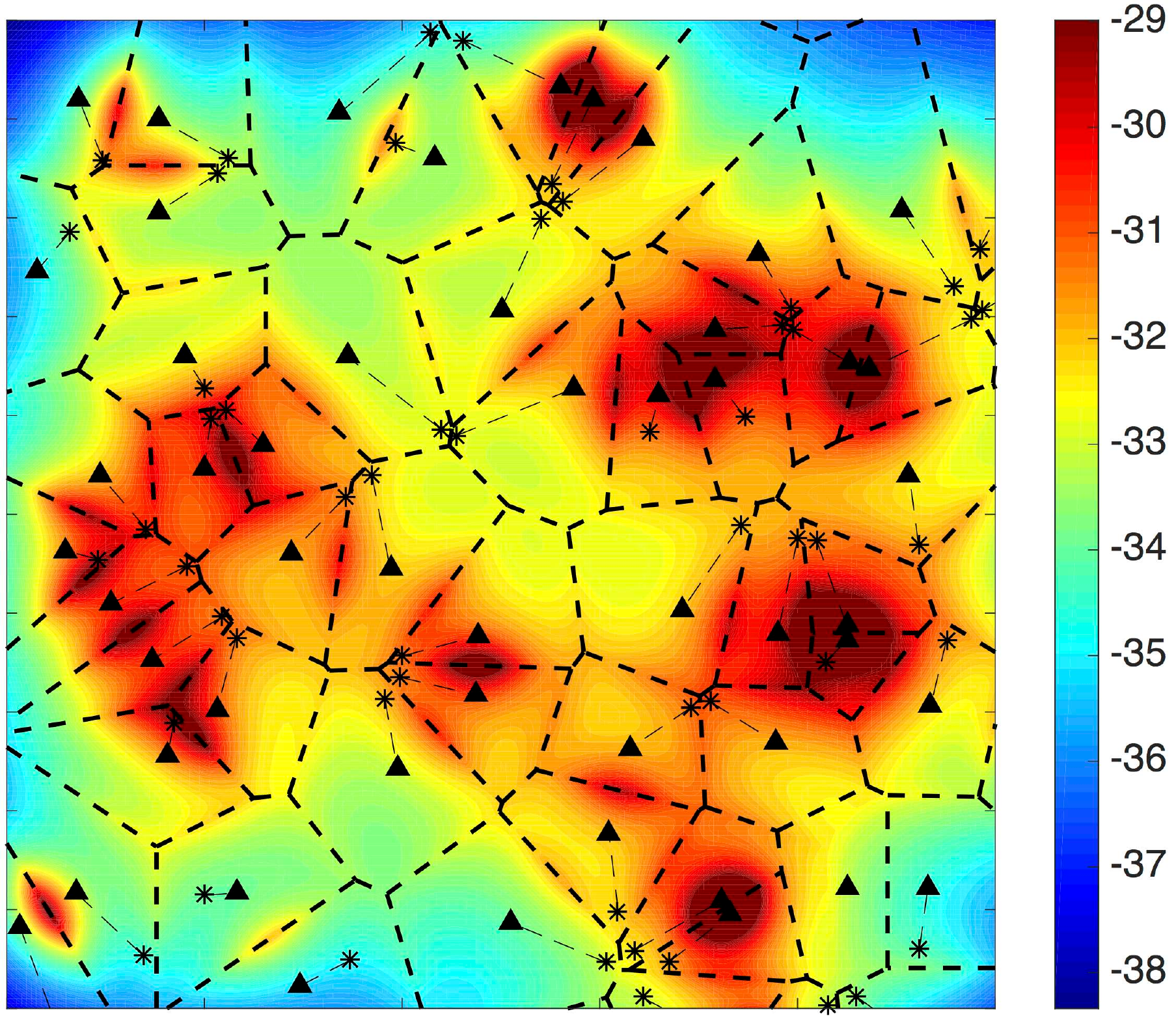}
	\caption{Downlink inter-cell interference}
        \label{fig:int_down}
	\end{subfigure}
	\begin{subfigure}[b]{0.32\textwidth}
	\includegraphics[width=\textwidth]{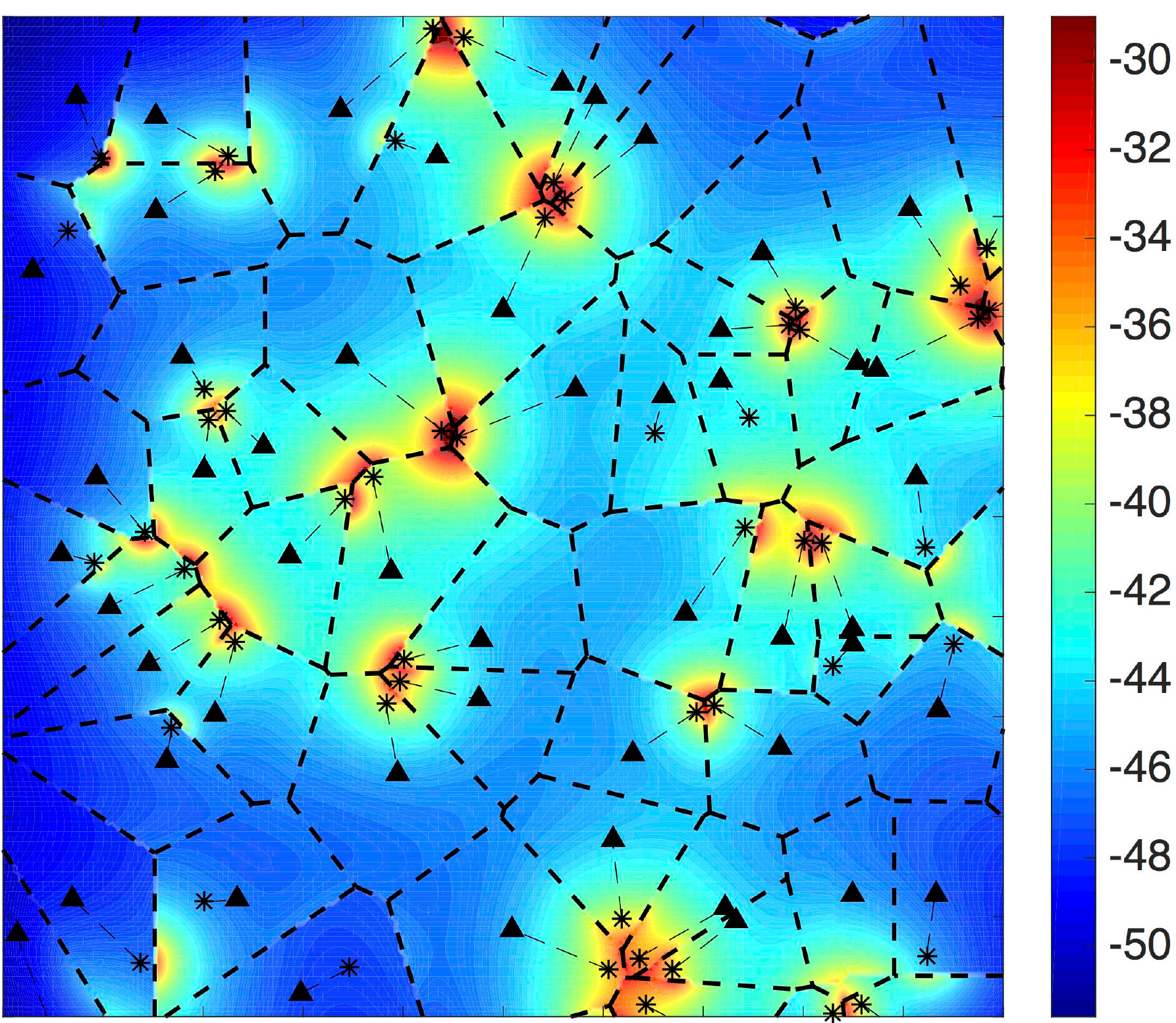}
	\caption{Uplink inter-cell interference}
        \label{fig:int_up}
	\end{subfigure}
		\begin{subfigure}[b]{0.32\textwidth}
	\includegraphics[width=\textwidth]{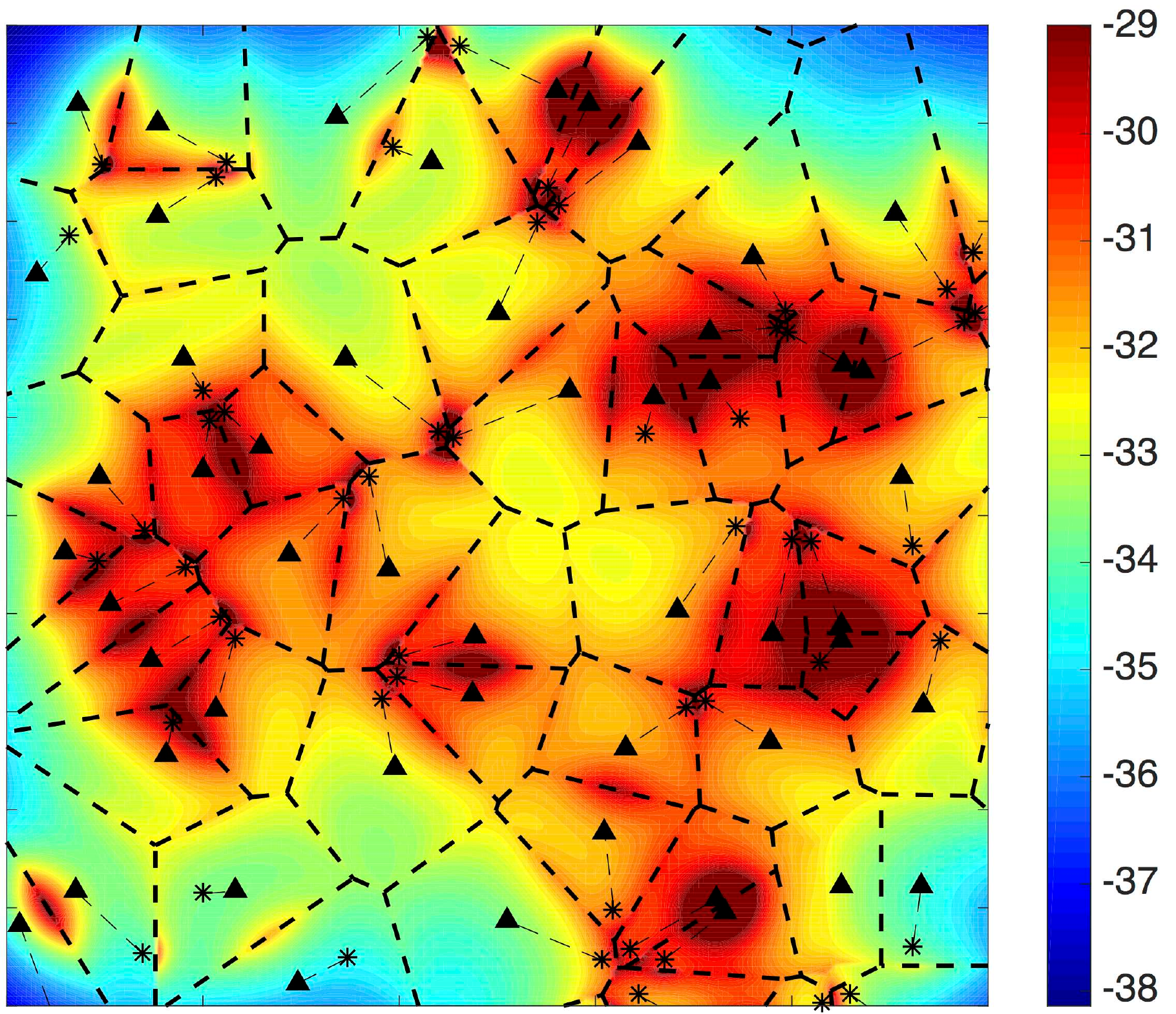}
	\caption{Full-duplex inter-cell interference }
        \label{fig:int_FD}
	\end{subfigure}
	\caption{ Aggregate inter-cell interference (in dBm) heat map for a cellular network of a major network operator in downtown London in which the BSs (represented by triangles) use constant transmit power of $P = 8$W, and UEs (represented by asterisks) employ a channel inversion power control with a target received power of $\rho = -50$dBm at the BSs.}
	\label{interference}
\end{figure*}

In the context of cellular networks, FD communication is optimistically promoted as a strong candidate to achieve the ambitious performance defined for 5G cellular networks\footnote{5G cellular networks are challenged to achieve 100-fold of the peak data rate and 1000-fold network capacity when compared to the state of the art 4G cellular networks \cite{what_5G}}~\cite{FD2}. However, SI is not the only obstacle to harvest the foreseen FD gains. FD communication increases the level of inter-cell interference, which imposes another major challenge for operating cellular networks in FD mode. Particularly, FD communication introduces uplink-to-downlink and downlink-to-uplink interference, hereafter denoted as cross-mode interference, because it allows base stations (BSs) and users equipment (UEs) to simultaneously transmit on the same set of channels that are reused over the spatial domain. It is worth mentioning that the conventional intra-mode inter-cell interference (i.e., downlink-to-downlink or uplink-to-uplink) is already a performance limiting parameter in modern cellular networks due to the tendency to use aggressive frequency reuse schemes (e.g., universal frequency reuse) and the deviation from regular grids to irregular topologies. Consequently, cross-mode interference challenges the success of FD communication in the cellular networks domain. 

In order to draw legitimate conclusions about FD enabled cellular networks, the effect of FD communication on the aggregate inter-cell interference and the subsequent effect on the overall network performance should be carefully studied. The authors in \cite{On2014Alves} show that despite the cross-mode interference induced by FD operation, the overall average spectral efficiency improves when compared to the HD operation. However, the study in \cite{On2014Alves} assume perfect SI cancellation. The effect of imperfect SI cancellation is studied in \cite{Hybrid2015Lee}, where the authors show that the FD gains in the downlink rate are proportional to the SI cancellation efficiency. While \cite{On2014Alves,Hybrid2015Lee} focus on the downlink performance, the authors in \cite{Analyzing2013Goyal} explicitly study the uplink and downlink performances in FD cellular networks. The study shows that uplink does not benefit from the FD operation and that the gains are mainly in the downlink direction. However, the study in \cite{Analyzing2013Goyal} is based on a simplistic system model that overlooks several aspects in the uplink operation that results in overestimating the uplink performance.  The authors in \cite{InBand2015Ahmad, Harvesting2016AlAmmouri, Limits2015Tsiky} show that the FD rate improvement in the downlink comes on the expense of significant degradation in the uplink rate. Consequently, the authors in \cite{InBand2015Ahmad, Harvesting2016AlAmmouri} propose the $\alpha$-duplex scheme which tunes the overlap between the uplink and downlink channels to balance the tradeoff between uplink and downlink rates. However, the results in \cite{InBand2015Ahmad, Harvesting2016AlAmmouri}  are based on a 2-D Poisson cellular network and simplistic propagation environment. 


This paper presents a study for FD operation in a realistic cellular environment, in which we explicitly account for uplink and downlink performances.  The realistic cellular environment is emulated via simulations with BSs locations drawn from a real cellular network in downtown London\footnote{BSs locations in the UK are available in Ofcom website: {http://sitefinder.ofcom.org.uk}} and propagation environment taken from the 3GPP recommendation~\cite{3gpp}. The study confirms that looking into the overall network performance may hide the uplink performance degradation and lead to misleading conclusions. In particular, the results show that the uplink is the bottleneck for FD operation due to the overwhelming cross-mode interference from BSs. Operating uplink and downlink on a common set of channels in an FD fashion improves the downlink rate but significantly degrade (over 1000-fold)  the uplink rate. Nevertheless, to exploit FD capabilities in the context of cellular networks, we employ the $\alpha-$duplex communication scheme, which allows partial overlap between uplink/downlink channels. The amount of overlap is controlled via the parameter $\alpha$, which is carefully tuned to increase the available bandwidth for uplink and downlink transmissions without sacrificing the uplink performance. Hence, simultaneous improvement in the uplink and downlink performances can be achieved. To this end, we discuss the backward compatibility of the $\alpha$-duplex scheme when implemented in BSs to serve HD users. Finally, we highlight potential research directions and conclude the paper.

\section{System Model}

We consider a $1000 \times1000$ m$^2$ urban area in downtown London with realistic BSs locations. The users are distributed according to a Poisson point process (PPP) over the same area. Average radio signal strength (RSS) based association is employed, which boils down to nearest BS association in the depicted system model. We focus on a single uplink/downlink channel pair which are universally reused by all BSs without intra-cell interference. Hence, only one active user is allowed in each cell. 
 
 All BSs transmit at a constant power level of $P_{\rm{d}}$ in the downlink. In the uplink, UEs employ the channel inversion power control scheme to compensate for the path-loss effect and maintain a target average power level of $\rho$ at the serving BS. Note that the employed channel inversion power control complies with the 3GPP recommendations for uplink transmission \cite{3gpp_power}. Due to the limited transmit powers of the UEs, we impose a maximum transmit power constraint of  $P^{({\rm{M}})}_{\rm{u}}$. Users that cannot maintain the required power level of $\rho$ transmit with their maximum power regardless of their locations.
 
We consider the 3GPP recommendation for path loss attenuation in urban environments \cite{3gpp}, in which the transmitted signals power attenuate with propagation distance according to
 \begin{equation}
 {\rm PL}(r) =22 \log(d)+ 28+20\log(f_c)
 \end{equation}
 where $d$ is the propagation distance in meters and $f_c$ is the carrier frequency in GHz. Beside path loss attenuation, transmitted signals experience Rayleigh fading such that the channel power gains are assumed to follow  i.i.d. exponential distribution with unit mean. 
 
 In this paper, we mainly asses the FD operation on the uplink and downlink performance by looking into their explicit ergodic rates. The ergodic rate is defined as:

\begin{align}
\mathcal{R} = \mathbb{E}\left[{\rm BW} \log\left(1+{\rm SINR}\right)\right],
\label{eq:rate}
\end{align}

\noindent where ${\rm BW}$ is the available bandwidth, ${\rm SINR}$ is the signal-to-interference-plus-noise-ratio at the decoder input of the intended receiver, and the expectation is over all ${\rm SINR}$ values. The ergodic rate measures the long-term maximum achievable rate and captures the positive and negative impacts of the FD operation in the $\rm BW$ term and the $\rm SINR$ term, respectively. It is worth noting that the average in \eqref{eq:rate} is calculated for the same realistic BSs locations but with different realizations for the users locations and channel gains.

 \section{FD Challenges and the $\alpha$-duplex Scheme} \label{duplex1}

The disparity in the transmit power between the BSs and UEs along with the irregular cellular structure are the main challenges for the FD operation in cellular networks. While the transmit power disparity leads to an overwhelming downlink to uplink interference, the irregular cellular structure increases the vulnerability of uplink transmissions to strong sources of interference. That is,  the RSS association fails to enforce a geographical interference protection in the uplink due to the irregular cell structure\cite{uplink_h, uplink_2}. To visualize the effect of the transmit power disparity in the FD operation we plot Fig.~\ref{interference}, which shows a real cellular network in downtown London along with the associated inter-cell interference heat-map assuming universal frequency reuse. The figure explicitly depicts the interference heat-map for downlink, uplink, and FD modes. The figure confirms the irregular structure of a real cellular network and highlights some cases where an uplink interferer can be closer to a BS than its intended uplink user. The figure also demonstrates that uplink transmissions reside in a less intense (12 dB less) inter-cell interference environment when compared to the downlink case. Consequently, the aggregated FD interference intensity depicted in Fig.~\ref{fig:int_FD} is comparable to the downlink interference shown in Fig.~\ref{fig:int_down}. On the other hand, comparing the interference intensity in Fig.~\ref{fig:int_up} and Fig.~\ref{fig:int_FD} manifests the challenge for uplink to survive in such extreme FD interference environment.

\begin{figure}[t]
    \centering
	\includegraphics[width=0.4\textwidth]{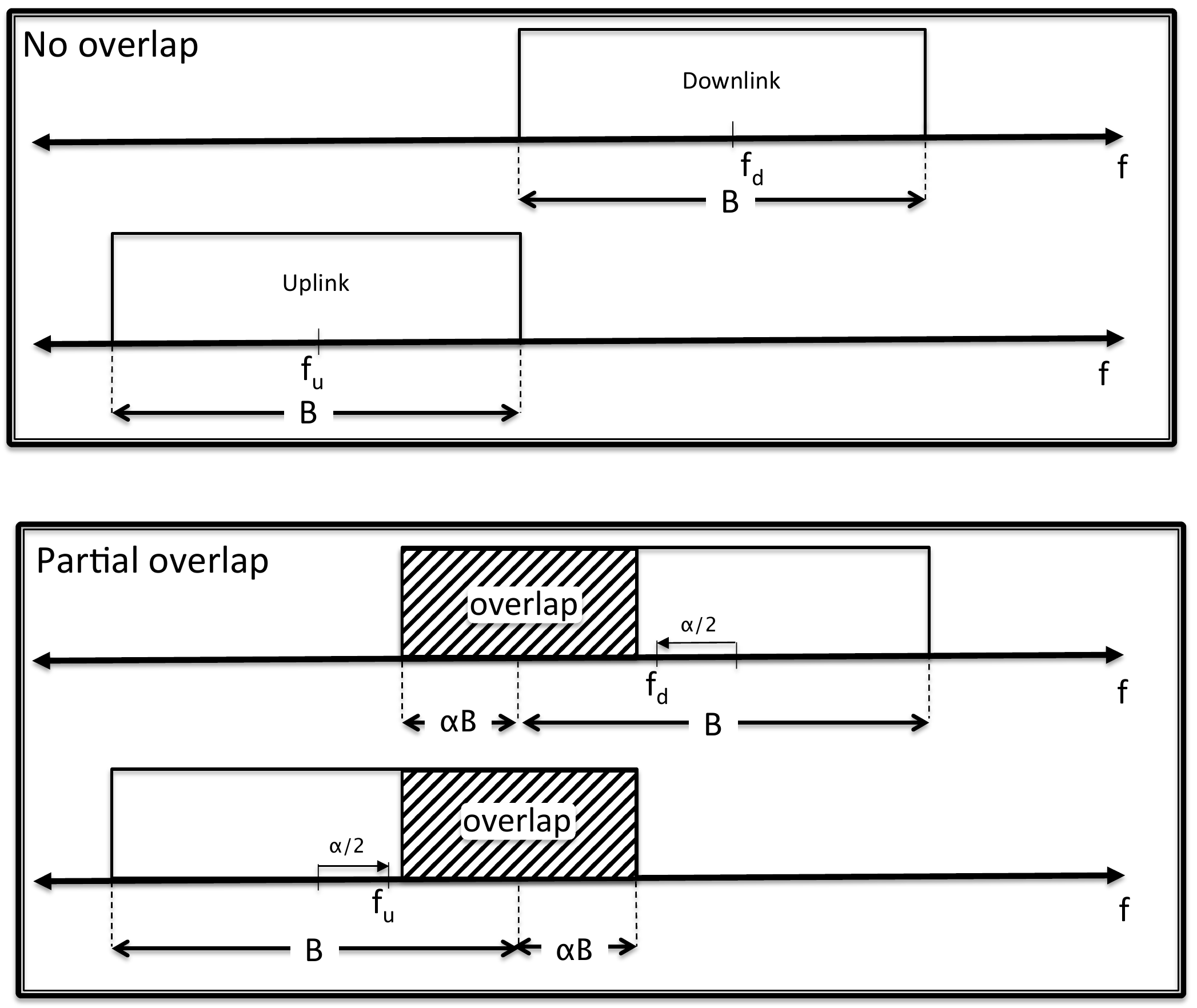}
	\caption{ Traditional half-duplex and proposed $\alpha$-duplex spectrum allocation for rectangular shaped power spectral density (PSD).}
	\label{duplex}
\end{figure}

Following \cite{InBand2015Ahmad}, we employ the $\alpha$-duplex scheme to balance the tradeoff between the uplink and downlink operation, which enables fine grained control for the overlap between uplink and downlink spectrum as shown in Fig.~\ref{duplex}.  Varying the duplexing parameter $ 0 \leq \alpha \leq 1$ over its support domain captures all possible duplexing schemes, in which the HD and FD special cases reside at the two extreme points $\alpha = 0$ and $\alpha = 1$, respectively.  The proposed $\alpha$-duplex scheme has practical and theoretical significance. From the practical point of view, tuning $\alpha$ can optimize the tradeoff between uplink and downlink rates. Note that, $\alpha$-duplex scheme also improves SI cancellation via an additional attenuation factor due to the misaligned uplink and downlink center frequencies.  From the theoretical point of view, tuning $\alpha$ helps characterizing the cross-mode interference by showing its gradual effect on the uplink and downlink rates. In order to have a thorough study for the $\alpha$-duplex scheme, we also consider the effect of pulse-shaping and matching filtering on signal-to-interference-plus-noise-ratio (SINR), and hence, on the performance. This is because the amount of cross-mode interference leaking into the decoder of the intended receiver is not only a function of the overlap parameter $\alpha$ but also depends on the pulse shaping templates used by the uplink and the downlink transmissions. 

\begin{figure*}[t]
	\centering
	\begin{subfigure}[b]{0.32\textwidth}
	\includegraphics[width=\textwidth]{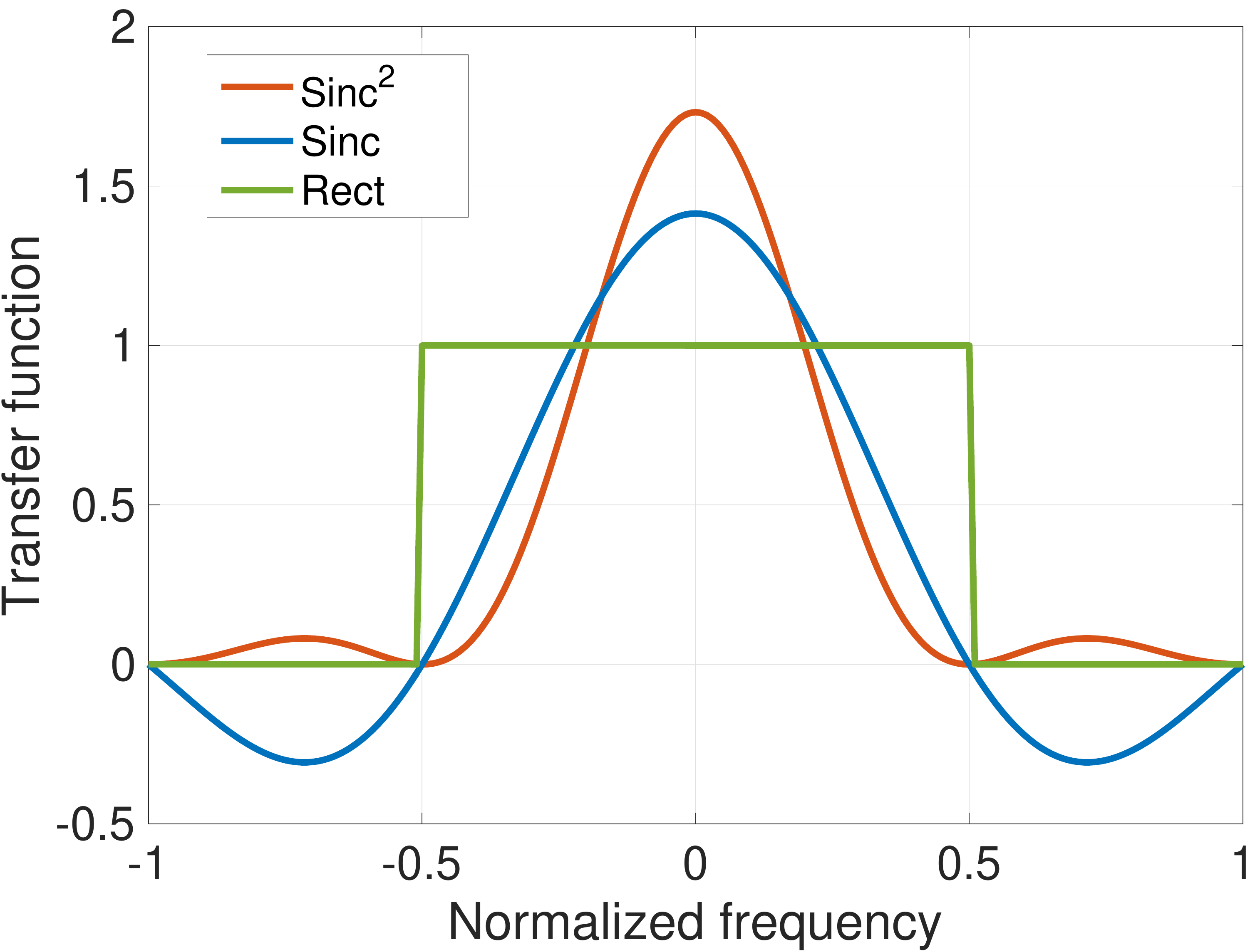}
	\caption{Normalized pulse shapes in the frequency domain.}
	        \label{fig:PS}
	\end{subfigure}
	\begin{subfigure}[b]{0.315\textwidth}
	\includegraphics[width=\textwidth]{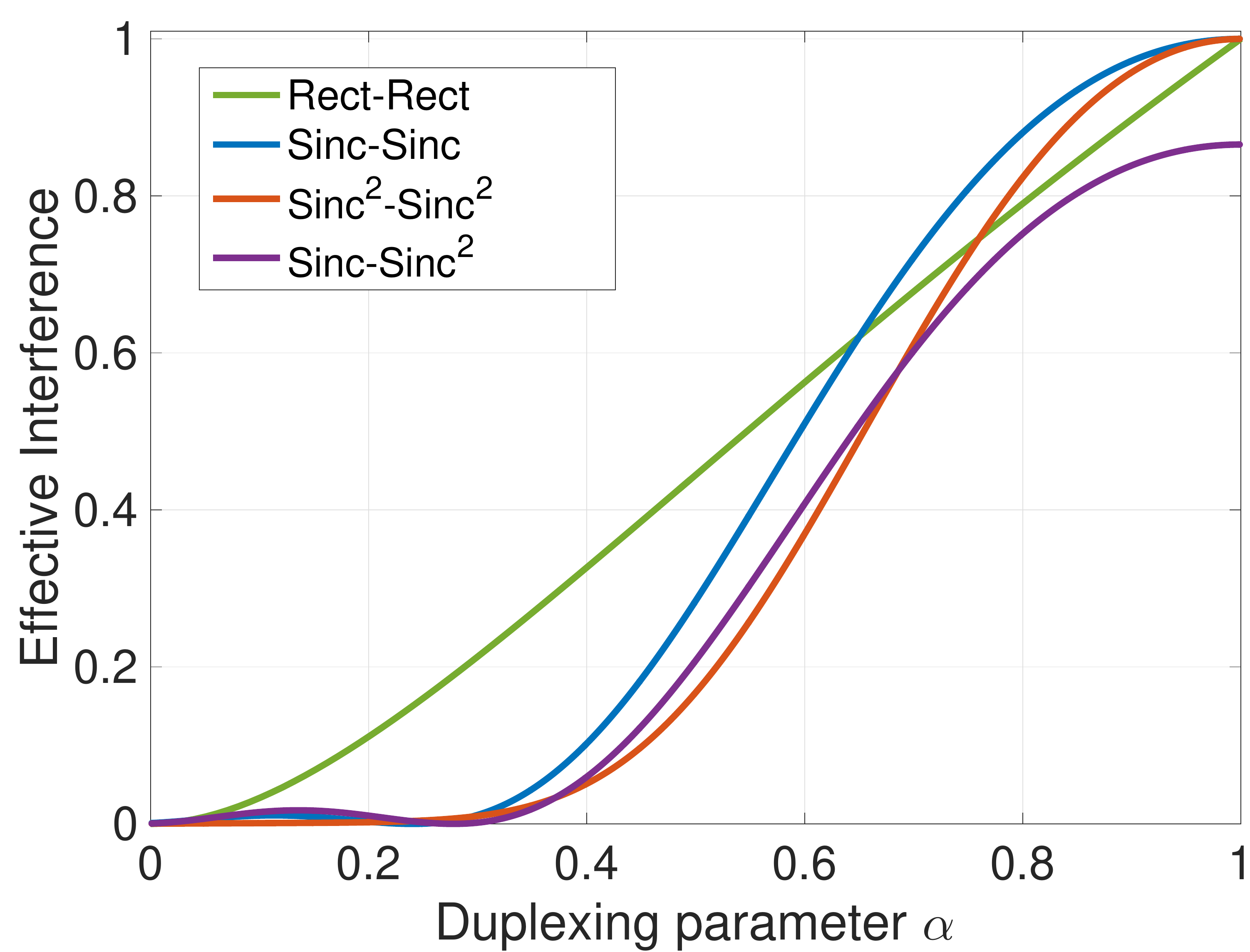}
	\caption{Effective interference vs the duplex parameter ($\alpha$).}
        \label{fig:EI}
	\end{subfigure}
		\begin{subfigure}[b]{0.315\textwidth}
	\includegraphics[width=\textwidth]{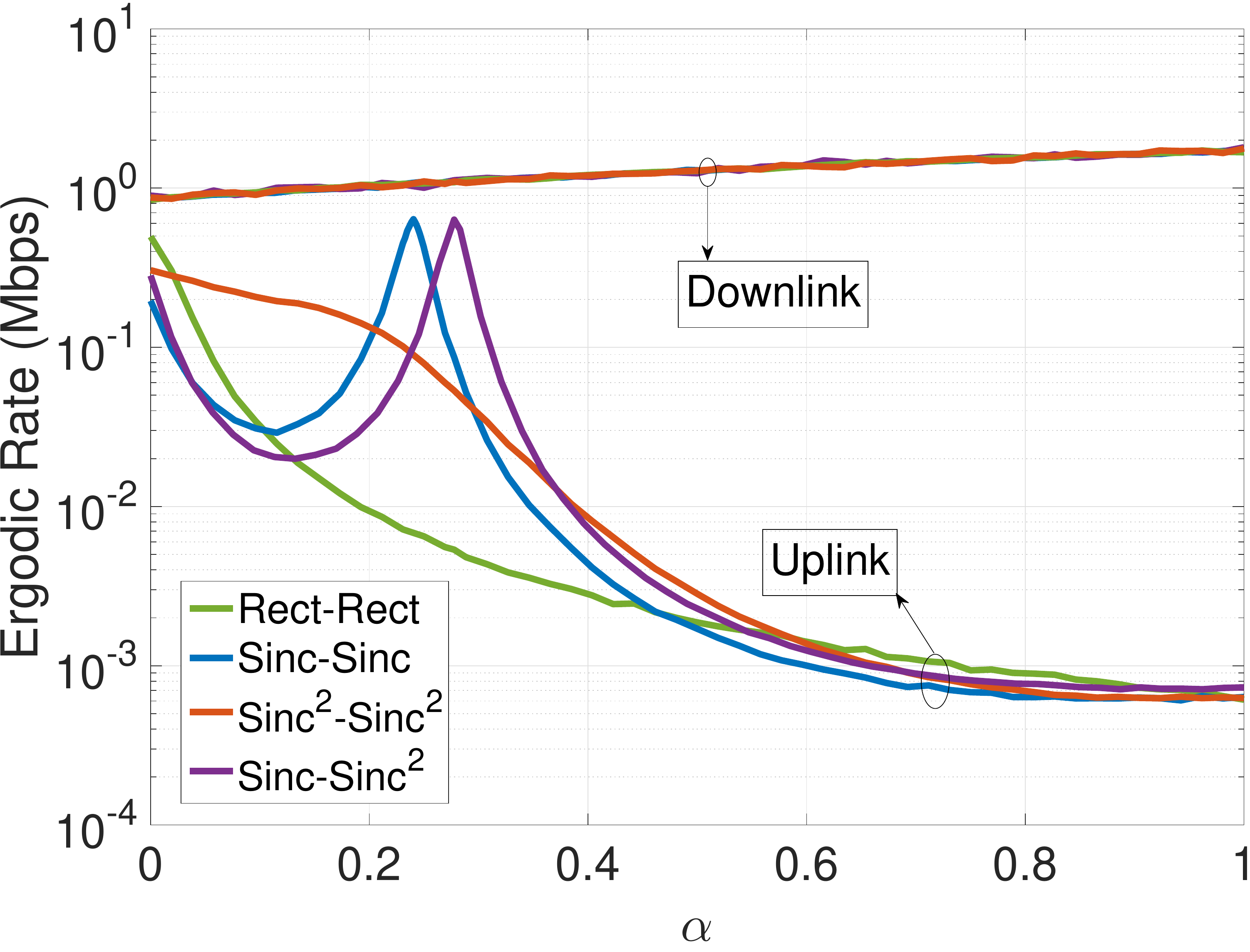}
	  \caption{Uplink and downlink rates vs the duplex parameter ($\alpha$) for different pulse shapes.}
	  \label{fig:per}	\end{subfigure}
	\caption{Pulse shapes, effective interference, and ergodic rate in the $\alpha$-duplex scheme.}
	\label{pulse_shape}
\end{figure*}

For simplicity, let us assume that the uplink and downlink have disjoint and adjacent channels of bandwidth $B$ Hz each, as shown in Fig.~\ref{duplex}. Furthermore, we assume a unified pulse shape for each mode of operation (i.e., uplink or downlink), however, the uplink and downlink do not necessarily use the same pulse shape. It should be mentioned that Fig.~\ref{duplex} shows only rectangular pulse shapes for the sake of illustration, however, the proposed system model employs different pulse shapes as shown in Fig.~\ref{fig:PS}. Increasing $\alpha$ simultaneously increases the ${\rm BW}$ of the uplink and downlink, which creates an overlap of $2 \alpha B$ between them. Allowing such overlap between uplink and downlink has two contradictory effects on the network rate. On one hand, the bandwidth ${\rm BW}$ available for the uplink and downlink transmissions increases at the rate of $(1+\alpha) B$,  in which the ergodic rate is linearly proportional to ${\rm BW}$. On the other hand, the cross-mode interference and SI leak into the receivers due to the overlap between uplink and downlink channels, which degrades the SINR. Note that the rate at which the SINR degrades with $\alpha$ depends on the used pulse shapes as discussed in the sequel.

At the receiver side, the matched filter convolves the received baseband (i.e., down-converted) signal with the conjugated time-reversed pulse shape template.  Then, the output of the matched filter is periodically sampled at the maximum peak of the output signal. The power at the output signal from the matched filter is proportional to the correlation between the signal pulse shape and the template used by the filter. Hence, cross-mode interference can be suppressed by selecting different pulse shapes for uplink and downlink. Furthermore, partial overlapping misaligns the uplink and downlink center frequencies, which creates a time offset for the peak of the cross-mode interference signal from the desired sampling point and further decreases the impact of the cross-mode interference. Consequently, cross-mode interference is multiplied by a factor of $\mathcal{E}(\alpha)$, denoted as the effective interference (EI) factor, to capture the matched filtering total effect. The EI factor is determined for a given $\alpha$ and known pulse shaping filters of both uplink and downlink based on the well-known concept of matched filtering   

\small
\begin{align}
{\cal E}(\alpha) \quad  {=} \int\nolimits_{-\frac{B+\alpha B}{2}}^{\frac{B+\alpha B}{2}} X(f- [1-\alpha]B) H^{*}(f) df,
\label{equ:Idd2}
\end{align}
\normalsize

\noindent where $X$ is the signal pulse shape in the frequency domain, and $H$ is the matched filter frequency response, and $(\cdot)^*$ denotes the conjugate operator. Fig.~\ref{fig:EI} shows the effect of the duplexing parameter $\alpha$ and the pulse shapes on the EI. The figure shows that the EI increases at different rates for different pulse shapes. A slowly increasing EI factor offers more bandwidth for the uplink and downlink transmissions with low cross-mode interference cost, which is desirable to improve the overall performance.

To incorporate the $\alpha$-duplex effect into the ergodic rate, \eqref{eq:rate} is rewritten as

\small
\begin{equation}\label{eq:rate2}
\!\!\mathcal{R}(\alpha)\! =\! \mathbb{E}\left[(1+\alpha) B \log\left(1+\frac{S}{\mathcal{I}+{\cal E}( \alpha) \left(\mathcal{C} + \beta \right) + \sigma^2(\alpha)}\right)\right]
\end{equation}
\normalsize

\begin{figure*}
\small
\begin{equation}\label{eq:UL}
\!\!\mathcal{R}_u(\alpha)\! =\! \mathbb{E}\left[(1+\alpha) B \log\left(1+\frac{P_{u_o} h_{u_o} L(\left \|b_o - u_o  \right \|)}{\underset{{u_i \in \Psi_u \setminus u_o}}{\sum} P_{u_i} h_{u_i} L(\left \|u_i - b_o  \right \|)+ {\cal E}( \alpha) \left(\underset{b_i \in \Psi_b \setminus b_o}{\sum} P_d h_{b_i} L(\left \|b_i - b_o  \right \|) + \beta \right) + \sigma^2(\alpha)}\right)\right]
\end{equation}
\normalsize
\hrulefill
\small
\begin{equation}\label{eq:DL}
\!\!\mathcal{R}_d(\alpha)\! =\! \mathbb{E}\left[(1+\alpha) B \log\left(1+\frac{P_{d} h_{b_o} L(\left \|b_o - u_o  \right \|)}{\underset{b_i  \in \Psi_b \setminus b_o}{\sum} P_d h_{b_i} L(\left \|b_i - u_o  \right \|)+{\cal E}( \alpha) \left(\underset{{u_i \in \Psi_u \setminus u_o}}{\sum} P_{u_i} h_{u_i} L(\left \|u_i - u_o  \right \|) + \beta \right) + \sigma^2(\alpha)}\right)\right]
\end{equation}
\normalsize
\hrulefill
\end{figure*}

\noindent where $S$ is the intended signal power, $\mathcal{I} $  is the intra-mode interference, $ \mathcal{C} $ is the cross-mode interference, and $\beta$ is the residual SI. Note that the intra-mode interference $ \mathcal{I} $  is not multiplied by an EI factor because transmissions in the same mode have perfectly aligned center-frequencies and use similar pulse shapes (i.e., EI is unity, see Fig.~\ref{fig:EI} at $\alpha=1$). On the other hand, SI is multiplied by EI because SI is a form of cross-mode interference but at the same transceiver. Furthermore, the noise power is a function of $\alpha$ because increasing the ${\rm BW}$ increases the noise leakage into the input of the decoder. In \eqref{eq:rate2} the expectation is over the random variables $S$, $\mathcal{I}$ and  $\mathcal{C}$. The explicit forms of \eqref{eq:rate2} for the uplink and downlink cases are given in \eqref{eq:UL} and \eqref{eq:DL}, respectively, where $\Psi_u$ is the set of all users locations, $\Psi_b$ is the set of all BSs locations, $u_o$ is the test user, $b_o$ is the test BS, $h_{x}$ is the channel gain between the test location and $x$, $\left \| \cdot \right \|$ is the Euclidean norm, and $L(\cdot) = 10^{-\frac{{\rm PL(\cdot)}}{10}}$ is the absolute value of path loss given in (1). It is worth mentioning that \eqref{eq:rate2},  \eqref{eq:UL}, and  \eqref{eq:DL} are unified for all duplexing schemes and captures the HD and FD as special cases at $\alpha=0$ and $\alpha=1$, respectively. 

The ergodic rate in \eqref{eq:UL} and \eqref{eq:DL}  are plotted in Fig.~\ref{fig:per} vs $\alpha$ for different combinations of pulse shapes and perfect SI cancellation at $P_d=5$W, $\rho=-70$dBm, and $B=1$ MHz. Note that the figure is plotted for perfect SI interference cancellation to emphasize the effect of cross-mode interference on the ergodic rate. The figure clearly shows the different behavior of the uplink and downlink with the duplexing parameter $\alpha$. While the downlink linearly increases in $\alpha$ for all pulse shapes, the uplink behavior is different and highly depends on the used pulse shapes. The linear increase of the downlink rate and the coincidence of the downlink curves confirm the negligible effect of the uplink-to-downlink interference (cf. Fig.~\ref{interference}). On the other hand, the different behavior of the uplink curves and the high degradation of the uplink rate at high values of $\alpha$ show the prominent impact of the downlink-to-uplink interference and the crucial influence of pulse shaping. Note that the behavior of each of the uplink curves in Fig.~\ref{fig:per} can be directly interpreted by looking into the EI factor at Fig.~\ref{fig:EI}. 

Looking into the numerical values, Fig.~\ref{fig:per} shows that operating in the FD mode (i.e., $\alpha=1$) almost doubles the downlink rate at the expense of more than 1000-fold degradation in the uplink rate. The figure also shows that using the $\alpha$-duplexing scheme and properly choosing the pulse shapes can result in a simultaneous improvement of $30\%$ in each of the uplink and downlink rates.

\section{Backward Compatibility of FD cellular Networks} \label{BC}

To harvest the aforementioned $\alpha$-duplex gains, FD BSs and FD UEs are required. However, cellular operators can only upgrade their BSs with FD transceivers and cannot enforce users to replace their HD terminal with FD terminals (i.e., with SI capabilities). Furthermore, SI cancellation can be too expensive to implement at the user side. Hence, backward compatibility is essential especially at the FD rollout phase.  

\begin{figure}[t]
	\centering
	\includegraphics[width=0.4\textwidth]{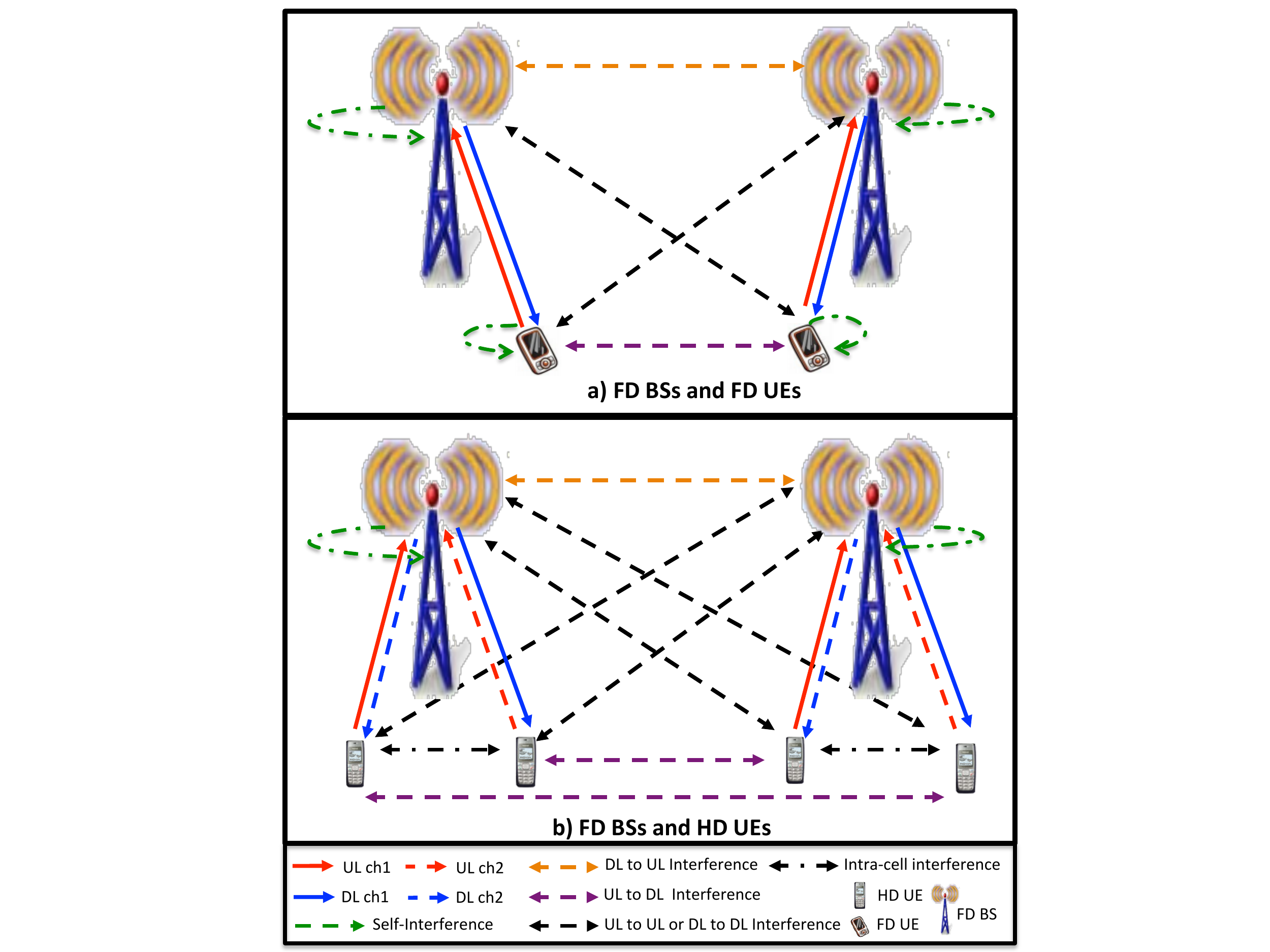}
	\caption{ Backward compatibility operation.}
	\label{back}
\end{figure}

HD UEs neither have the technological capability to simultaneously transmit and receive on the same channel nor can tolerate SI. Nevertheless, FD transceivers at the BSs can still be exploited to serve several HD UEs on mutually overlapped channels as shown in Fig.~\ref{back}. Similar to the $\alpha$-duplex scheme introduced in Section~\ref{duplex1}, uplink and downlink channel pairs are partially overlapped, however, each UE is assigned uplink and downlink channels from different channel pairs (cf. Fig.~\ref{back}).  Consequently, an overlap free assignment is enforced at the user side, which eliminates UEs' SI at the expense of introducing mutual cross-mode intra-cell interference. In other words, in the backward compatibility mode, each downlink transmission has a single cross-mode intra-cell interferer. Such intra-mode interference can be suppressed by exploiting multi-user diversity in which UEs with poor mutual channel conditions are coupled on the same channels. At the BS side, the backward compatibility mode does not eliminate SI because each BS transmits and receives on all channel pairs that are assigned to its users. 

\begin{figure}[t]
\centerline{\includegraphics[width=  0.4\textwidth]{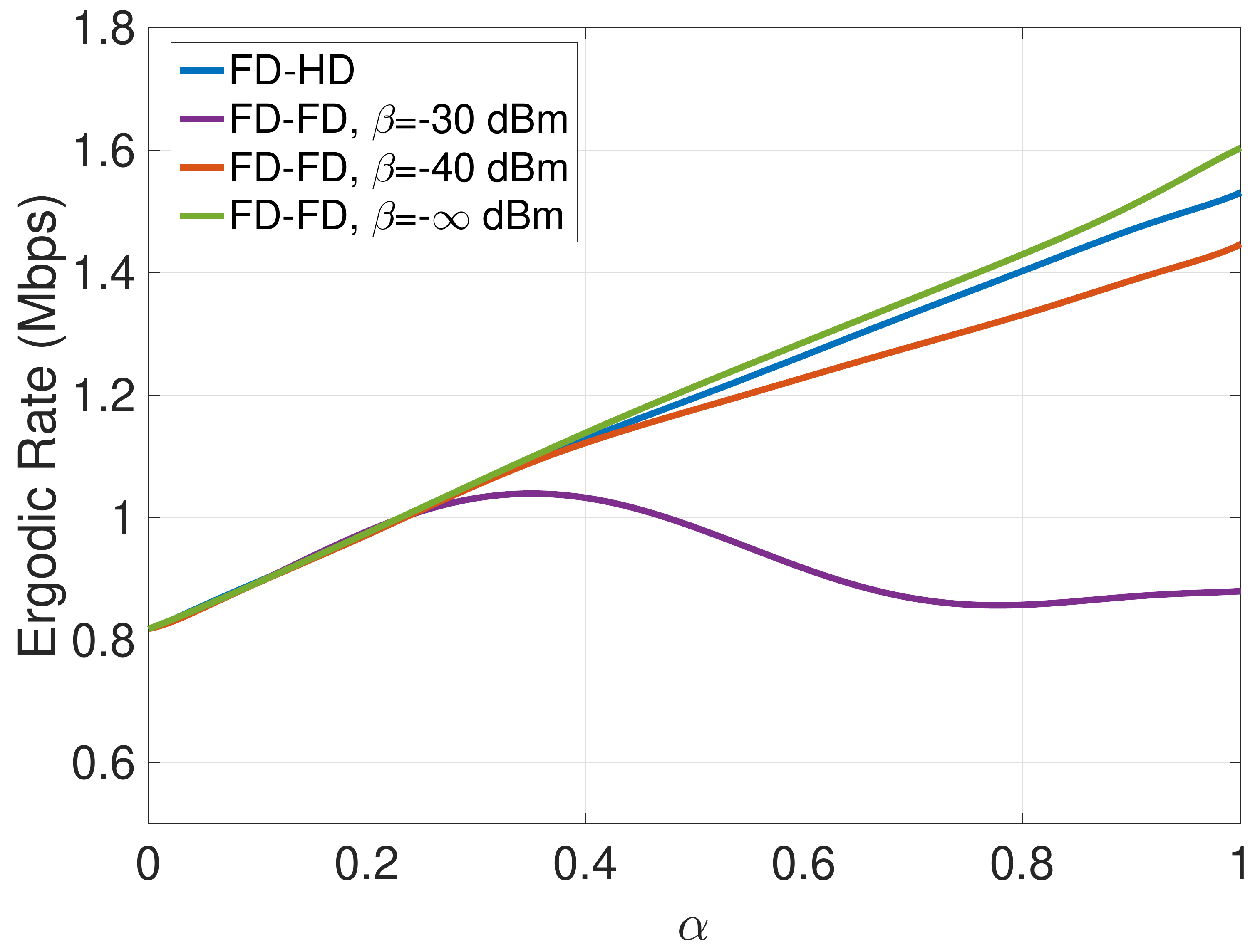}}\caption{\,Downlink rate vs $\alpha$ for HD UEs and FD UEs with different values of $\beta$.}
\label{fig:back}
\end{figure}

Fig.~\ref{fig:back} shows the downlink rate for the $\alpha$-duplex scheme with FD UEs that have different SI cancelation capabilities as well as for the  $\alpha$-duplex scheme with HD UEs. The figure shows a slight loss in the downlink rate (at most $5\%$) occurs at high values of $\alpha$ for HD UEs when compared to FD UEs that have high SI cancellation efficiency {( $\beta \leq -40 $ dBm)}. On the other hand, the downlink rate for HD UEs outperforms that of the FD UEs that have poor SI cancellation efficiency, especially at high values of $\alpha$. Consequently, it can be concluded that FD receivers are not mandatory to harvest the $\alpha$-duplex gains. Another interesting conclusion from Fig.~\ref{fig:back} is that the backward compatibility mode should be the default mode of operation if the SI cancellation efficiency of the UEs are unknown. This would avoid high performance degradation in case that the UEs are not equipped with efficient FD transceivers. It is worth mentioning that backward compatibility mode is proposed in \cite{Full2014Sundaresan, Hybrid2015Lima} for FD only, which overlooks the uplink rate degradation problem shown in Fig.~\ref{fig:per}. 

The uplink rate in the backward compatibility mode is identical to Fig.~\ref{fig:per}, and hence, is not shown in Fig.~\ref{fig:back}. Consequently, the $\alpha$-duplex design introduced in Section~\ref{duplex1} is also necessary for the backward compatibility mode to avoid significant degradation in the uplink rate. Having said that, it should be noted that the downlink rate for HD UEs mode is almost identical to that of the FD UEs with perfect SI cancellation when operating at low values of $\alpha$. 

It is worth mentioning that the behaviors shown in  Fig.~\ref{pulse_shape} and Fig.~\ref{fig:back} are consistent with our finding in \cite{InBand2015Ahmad, Harvesting2016AlAmmouri} that are based on stochastic geometry  analysis rather than the presented realistic simulation. Nevertheless, the numeric values obtained in  Fig.~\ref{pulse_shape} and Fig.~\ref{fig:back} are different and Fig.~\ref{pulse_shape} shows more significant uplink degradation (100-fold less rate than \cite{InBand2015Ahmad}) in the uplink at the FD operation. While the consistency between the behaviors in the presented study and the results in \cite{InBand2015Ahmad, Harvesting2016AlAmmouri} proves the generality of the findings in this paper, the obtained numeric values manifests the uplink vulnerability in the FD mode.

 \section{Future Research Directions} \label{fut}

FD duplex operation introduces cross-mode interference, which is a new and non-trivial performance limiting parameter for cellular networks. As shown in this paper, cross-mode interference has its own unique characteristics that differ from the conventional HD interference. Consequently, cellular network functions that are tailored to mitigate HD interference only should be revisited and adapted to deal also with cross-mode interference. Furthermore, new interference management techniques may be required to further mitigate the cross-mode interference effects. This section highlights some future research directions in the context of FD enabled cellular networks.

\begin{itemize}


\item \textbf{Pulse shaping and matched filtering:} This study considers well-known pulse shaping templates to mitigate cross-mode interference leakage into decoders in the $\alpha$-duplex scheme. A more efficient way is to customize pulse shapes that are optimized to achieve the lowest EI factor at the highest possible overlap. Such problem can be formulated by following a similar analogy to the one presented in \cite{Siala}. Note that the matched filtering optimizer in the $\alpha$-duplex scheme is different from the conventional case as the main objective in the $\alpha$-duplex case is to suppress the cross-mode interference in order to maximize the SINR.

\item \textbf{Distributed duplexing and pulse shaping:} It is not mandatory to select a system wide duplexing parameter $\alpha$ and unified pulse shapes for the uplink and downlink. A more interesting and general case is to consider per user pulse shaping and per-cell duplexing. Distributed duplexing and pulse shaping can be realized by the aid of context-aware cloud assisted networking that assigns pulse shapes and duplexing parameters in addition to the other network resources (e.g., power and channels) to the BSs and UEs. That is, depending on the users and BSs relative locations, channel gains, and SINR feedback, an optimal per user duplexing can be obtained.

\item \textbf{Interference management:} Cross-mode interference mitigation via the $\alpha$-duplexing scheme can be insufficient to meet the desired quality of service constraints for error sensitive applications. Hence, interference management techniques such as beamforming, interference alignment, BSs cooperation, power control should be integrated on top of the $\alpha$-duplexing scheme to further reduce cross-mode interference are required. 

\item \textbf{Multi-tier architecture:} Modern cellular networks are constituted from different types of BSs with different properties (i.e., transmit power, intensities, number of antennas, etc.). Hence, the basic RSS association is replaced with more sophisticated association techniques that may enforce a decoupled uplink/downlink association~\cite{uplink_2}. Different BS association along with FD operation creates interference between the downlink and uplink transmissions of the same UE that cannot be mitigated via SI cancellation. Hence, multi-tier cross-mode interference mitigation techniques are required.   


\end{itemize}
 
\section{Conclusion} 

This paper  presents a study on FD enabled cellular networks in a realistic cellular environment. The study manifests the challenge for uplink transmissions to operate in FD cellular environments due to the overwhelming downlink interference. In particular, the results show that FD operation almost doubles the downlink rate at the expense of more than 1000-fold degradation is the uplink rate. Therefore, we propose the $\alpha$-duplex scheme that relies on partial spectrum overlapping between uplink and downlink along with pulse shaping and matched filtering to maintain an acceptable uplink performance in FD cellular environments. The results confirm the superiority of the $\alpha$-duplex scheme over both the FD and HD schemes, in which a simultaneous improvement of $30\%$ for each of the uplink and downlink rates are achieved. This paper  also demonstrates the backward compatibility of the $\alpha$-duplex scheme and shows that FD user terminals are not necessary to harvest the $\alpha$-duplex gains. To this end, we highlight future research directions for FD communication in 5G cellular networks.

\bibliographystyle{IEEEtran}
\bibliography{IEEEabrv,reference}

\vfill

\end{document}